\documentclass[letterpaper,twocolumn,fleqn]{article} 

\usepackage{ist}
\usepackage{booktabs}
\usepackage{amsmath}
\usepackage{amsfonts}
\usepackage{xspace}
\usepackage[noend]{algpseudocode}
\usepackage[ruled,vlined,algo2e,linesnumbered]{algorithm2e}
\usepackage[group-separator={,}, group-minimum-digits={3}]{siunitx}
\usepackage{subfig}
\usepackage{url}
\usepackage[colorlinks = true,
            linkcolor = blue,
            urlcolor  = blue,
            citecolor = blue,
            anchorcolor = blue]{hyperref}

\title{Digital Reconstruction of Elmina Castle for Mobile Virtual Reality via Point-based Detail Transfer}

\author{Sifan Ye, Computer Science Department, Stanford University, Palo Alto, CA  USA\\
\textit{Ting Wu}, eBay, San Jose, CA USA\\
\textit{Michael Jarvis}, Department of History,  University of Rochester, Rochester, NY USA\\
\textit{Yuhao Zhu}, Department of Computer Science,  University of Rochester, Rochester, NY USA
}

\date{} 

\begin{document} 


\newcommand{\website}[1]{{\tt #1}}
\newcommand{\program}[1]{{\tt #1}}
\newcommand{\benchmark}[1]{{\it #1}}
\newcommand{\fixme}[1]{{\textcolor{red}{\textit{#1}}}}

\newcommand*\circled[2]{\tikz[baseline=(char.base)]{
            \node[shape=circle,fill=black,inner sep=1pt] (char) {\textcolor{#1}{{\footnotesize #2}}};}}

\ifx\figurename\undefined \def\figurename{Figure}\fi
\renewcommand{\figurename}{Fig.}
\renewcommand{\paragraph}[1]{\textbf{#1} }
\newcommand{\figline}{{\vspace*{.05in}\hline}}
\newcommand{\funcname}[1]{\textsc{\textbf{\textcolor{blue}{#1}}}}

\newcommand{\Sect}[1]{Section~\ref{#1}}
\newcommand{\Fig}[1]{Figure~\ref{#1}}
\newcommand{\Tbl}[1]{Table.~\ref{#1}}
\newcommand{\Equ}[1]{Equ.~\ref{#1}}
\newcommand{\Apx}[1]{Apdx.~\ref{#1}}
\newcommand{\Alg}[1]{Algorithm~\ref{#1}}

\newcommand{\specialcell}[2][c]{\begin{tabular}[#1]{@{}c@{}}#2\end{tabular}}
\newcommand{\note}[1]{\textcolor{red}{#1}}

\newcommand{\hpm}{\textsc{HPM}\xspace}
\newcommand{\lpm}{\textsc{LPM}\xspace}
\newcommand{\trans}{\textsc{Remeshing}\xspace}

\newcommand{\proj}{\textsc{Mesorasi}\xspace}
\newcommand{\mode}[1]{\underline{\textsc{#1}}\xspace}
\newcommand{\sys}[1]{\underline{\textsc{#1}}}

\newcommand{\RNum}[1]{\uppercase\expandafter{\romannumeral #1\relax}}

\def\cA{{\mathcal{A}}}
\def\cF{{\mathcal{F}}}
\def\cN{{\mathcal{N}}}


\graphicspath{{figs/}}

\maketitle 

\thispagestyle{empty} 


\begin{abstract}
Reconstructing 3D models from large, dense point clouds is critical to enable Virtual Reality (VR) as a platform for entertainment, education, and heritage preservation. Existing 3D reconstruction systems inevitably make trade-offs between three conflicting goals: the efficiency of reconstruction (e.g., time and memory requirements), the visual quality of the constructed scene, and the rendering speed on the VR device. This paper proposes a reconstruction system that simultaneously meets all three goals. The key idea is to avoid the resource-demanding process of reconstructing a high-polygon mesh altogether. Instead, we propose to directly transfer details from the original point cloud to a low polygon mesh, which significantly reduces the reconstruction time and cost, preserves the scene details, and enables real-time rendering on mobile VR devices.

While our technique is general, we demonstrate it in reconstructing cultural heritage sites. We for the first time digitally reconstruct the Elmina Castle, a UNESCO world heritage site at Ghana, from billions of laser-scanned points. The reconstruction process executes on low-end desktop systems without requiring high processing power, making it accessible to the broad community. The reconstructed scenes render on Oculus Go in 60 FPS, providing a real-time VR experience with high visual quality.
\end{abstract}

\section{Introduction}
\label{sec:intro}

Historic sites of cultural significance are being destroyed on a daily basis. While it is impossible to fully prevent this, we can digitally preserve important historic sites by reconstructing these sites in 3D before destruction for later research. Apart from academic research, reconstructing cultural heritage sites is also instrumental to educating and entertaining the general public; websites such as CyArk~\cite{cyark}, Open Heritage 3D~\cite{openheritage}, and Bermuda 100~\cite{bermuda100} promise to let people virtually visit ``the world's most famous monuments in immersive and accurate 3D.'' Humanities and education researchers have long advocated using digital interactive games/simulations to transport students across time and space to better understand the past and other cultures~\cite{gee2003video, champion2003applying, kapell2013playing, squire2011video}.

This paper describes our effort to reconstruct Elmina Castle, a historical slave trade castle and a UNESCO World Heritage Site in Ghana, from large-scale laser-scanned point clouds, which we collected during field trips to Ghana over the past three years. We target mobile Virtual Reality (VR) headsets, e.g., Oculus Go, which provide an ideal platform for immersive experience but also present the challenge of limited processing power. 

Besides the historical and humanitarian significance of the reconstructed site itself, we present a reconstruction workflow that addresses two technical challenges: 1) the reconstruction algorithm must generate 3D models that can be rendered in real-time on resource-limited mobile VR devices with high quality, and 2) the reconstruction algorithm itself must be lightweight so as to be accessible to the broad archaeology community. Let us elaborate.

First, rendering large-scale 3D models on mobile VR headsets is challenging due to the limited amount of compute resources. A 3D model of even a single room in Elmina Castle has millions of polygons, which render at merely 10 FPS on Oculus Go. Existing workflows either simply use a low-polygon mesh or decimate a high-polygon mesh~\cite{cohen1998appearance, garland1997surface} (i.e., remeshing). Both yield smaller meshes and improve the rendering speed but usually come at a cost of losing critical scene details, which is disruptive when closely examining an artifact.



Second, we also aim to develop a lightweight reconstruction workflow that makes reconstructing heritage sites from huge point clouds more accessible to historians and archaeologists, who may not have access to powerful computers and could not afford long reconstruction time. For instance, the entire scan of Elmina Castle contains 33 billions of points; reconstruct the castle on a high-end Intel Xeon server with 256 GB RAM and 2 Nvidia GTX 1080Ti GPUs takes three weeks, which in reality is even longer because the constructed mesh could be iteratively edited.

Our reconstruction workflow simultaneously addresses the two challenges above. We completely skip high-polygon meshes altogether. Instead, we use a low-polygon mesh as the base representation of the scene, and then transfers the details, e.g., color and normal, \textit{directly} from the original point cloud to the low-polygon mesh. Our reconstruction workflow is lightweight and fast since we never reconstruct a high-polygon mesh throughout the whole process. The reconstructed model can be rendered in real-time since the underlying representation is a low-polygon mesh, which, critically, is associated with high-quality texture information transferred from the original point clouds.

\begin{figure*}[t]
  \begin{minipage}[t]{1\columnwidth}
    \centering
    \includegraphics[trim=0 0 0 0, clip, height=2in]{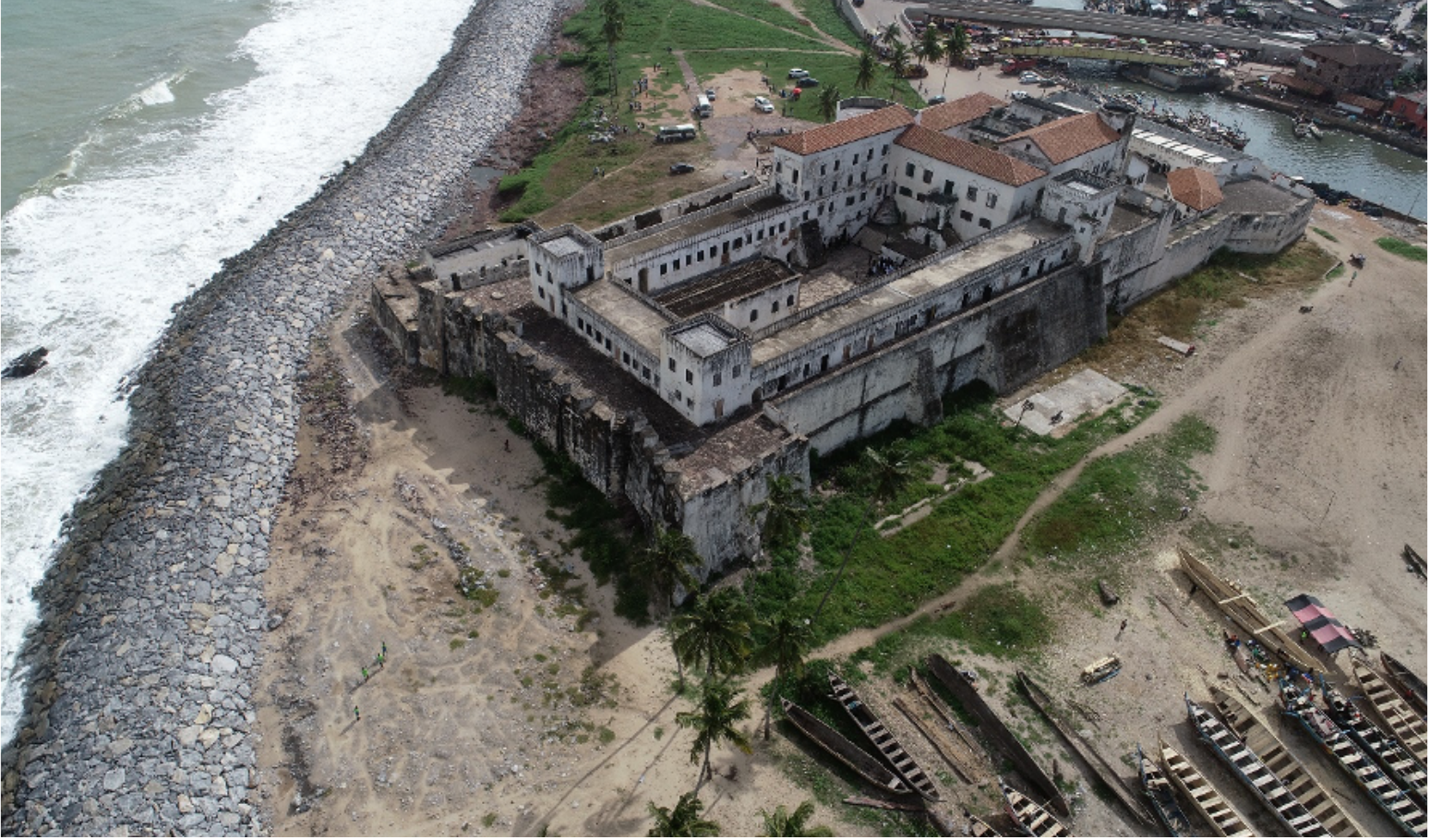}
    \caption{Overlook of Elmina Castle.}
    \label{fig:castle}
  \end{minipage}
  \hfill
  \begin{minipage}[t]{1\columnwidth}
    \centering
    \includegraphics[trim=0 0 0 0, clip, height=2in]{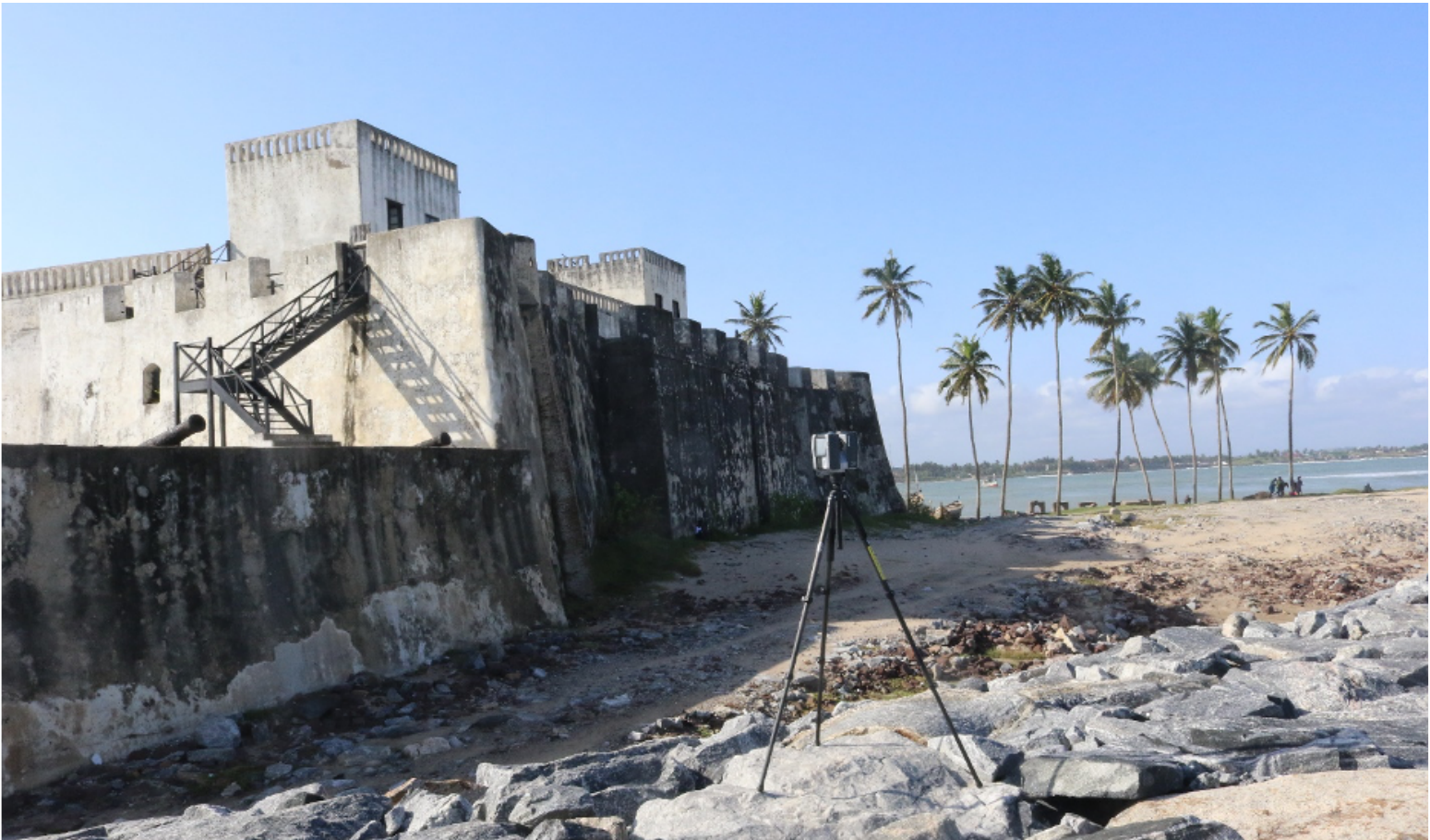}
    \caption{Scanning the exterior of the castle.}
    \label{fig:scanner}
  \end{minipage}
\end{figure*}

Overall, our reconstruction system achieves a wall-clock speedup of up to 3.1 times compared with an existing commercial reconstruction workflow. On average, our reconstruction system converts scans with over 30 million points to meshes with 200 thousand polygons, which render in 60 FPS on Oculus Go, a representative mobile VR headset, while presenting a desirable visual quality, evaluated both objectively and subjectively.

To encourage further digitization of the castle and provide immersive learning, research, and education experience to people who could not visit the castle in person, our system is available at \url{https://github.com/horizon-research/3D-Reconstruction-From-Point-Cloud}.

%


\section{Context and Background}
\label{sec:bck}


\subsection{Why Digitally Reconstructing Elmina Castle?}
\label{sec:bck:elmina}

\paragraph{Historical Significance} Built in 1482 by the Portuguese, Elmina Castle (\Fig{fig:castle}) is the first European trading base constructed in Sub-Saharan Africa. For more than five hundred years, it was a commercial hub where European goods were exchanged for gold, ivory, and slaves. During the Portuguese period (1482-1637) and after the Dutch West India Company captured it in 1637, Elmina Castle was the central administrative base for commercial operations at satellite forts stretching hundreds along the Gold and Slave Coasts. For their central role in the trans-Atlantic slave trade and as a unique collection of European fortifications adapted to an African environment, Ghana’s slave trade forts and castles were given UNESCO World Heritage Site status in 1979.

\paragraph{Humanity} Elmina holds personal relevance for tens of millions of people globally. Elmina Castle is a major center for directing the African slave trade and the oldest site of sustained Euro-African contact and commerce in Sub-Saharan Africa. Under the Ghana Museums and Monuments Board, the well-preserved site (along with nearby Cape Coast Castle) has become a major pilgrimage site of African Diaspora heritage tourism, attracting tens of thousands of African Americans annually seeking insights into their ancestors' experiences of enslavement. Creating a virtual experience through a digital surrogate has immense humanity value for those who could not physically visit the castle, whose importance is only heightened by COVID-19.

\paragraph{Research and Education} Virtually reconstructing Elmina has significant research value for archaeologists, historians, and mechanical engineers. For instance, archaeologists use the reconstructed fabric of the castle's floors, walls, and ceilings to date the castle's various rooms and determine the structure's evolution. Early Portuguese parts of the castle would be indicated by the use of mud mortar, thin red bricks, and roof tiles imported from Portugal. Later Dutch repairs and expansions of the castle incorporated lime mortal, imported yellow Dutch bricks, and angled ramparts designed to sustain cannon fire. Mechanical engineers could view the reconstructed model to dynamically model the monument's structural  integrity and recommend repairs.

\subsection{Scanning System}
\label{sec:bck:pc}

Over a three-year span, we use a FARO Focus3D X 130 laser scanner~\cite{farox130} to generate colored point clouds for each of the castle's 120 rooms and exterior areas (\Fig{fig:scanner}). For each scan, the time-of-flight ranging unit first generates a raw (colorless) point cloud of the scene; on a separate pass, a color camera built-in with the scanner captures the color images, which are then later registered with the raw point cloud to generate a colored point cloud.

Overlapped individual scans within rooms were first registered using FARO's Scene software; the outliers were then removed to create the final dense point cloud, which covers every room. By combining a total of 427 individual scans, the point cloud for the entire castle exceeds 33 billion points with an average resolution (distance between neighboring points) of \SI{1.9}{\mm}. In tests conducted offline, our scans have a point accuracy of \SI{4.7}{\mm} overall; most rooms have point accuracies under \SI{2.0}{\mm}.


\begin{figure*}[t]
    \centering
    \includegraphics[width=2\columnwidth]{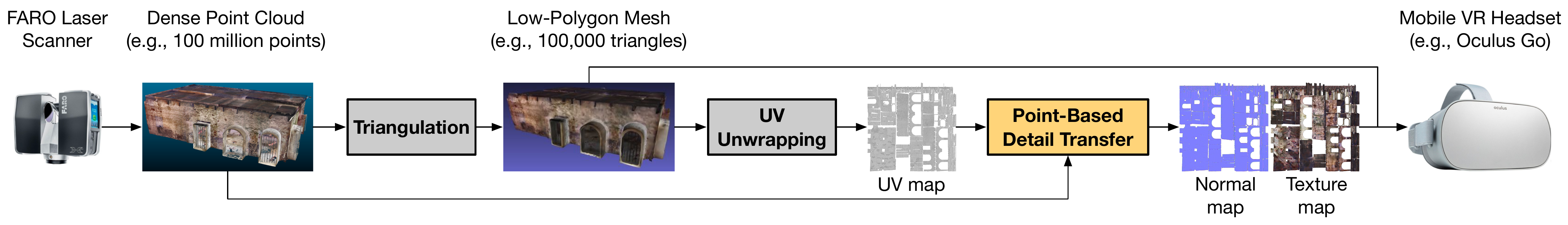}
    \caption{End-to-end reconstruction pipeline from the dense laser scanned point to mobile VR rendering. We transfer scene details directly from the point cloud to the UV map generated from a low-polygon mesh. The low-polygon mesh ensures high rendering speed on mobile VR headsets while the detail transfer ensures satisfactory visual quality.}
    \label{fig:pipeline}
\end{figure*}

\subsection{Point Cloud Reconstruction for Mobile VR}
\label{sec:bck:reconstruct}

While it is possible to directly render the point clouds~\cite{schutz2019real, bonatto2016explorations}, doing so has two key disadvantages. First, pure point clouds are just not visually appealing. Users can see ``holes'' in the scenes, which is particularly disruptive when users want to closely examine a historical artifact. Second, many VR use-cases such as virtual tours allows users to interact with the scene, e.g., walking around the castle, or observing avatars of historical figures living/working in the castle, which in turn requires a continuous surface to simulate collision for realistic interactions. Thus, we reconstruct polygon meshes from the colored point clouds.

\section{Design Objectives}
\label{sec:cha}

\begin{figure}[t]
    \centering
    \includegraphics[width=\columnwidth]{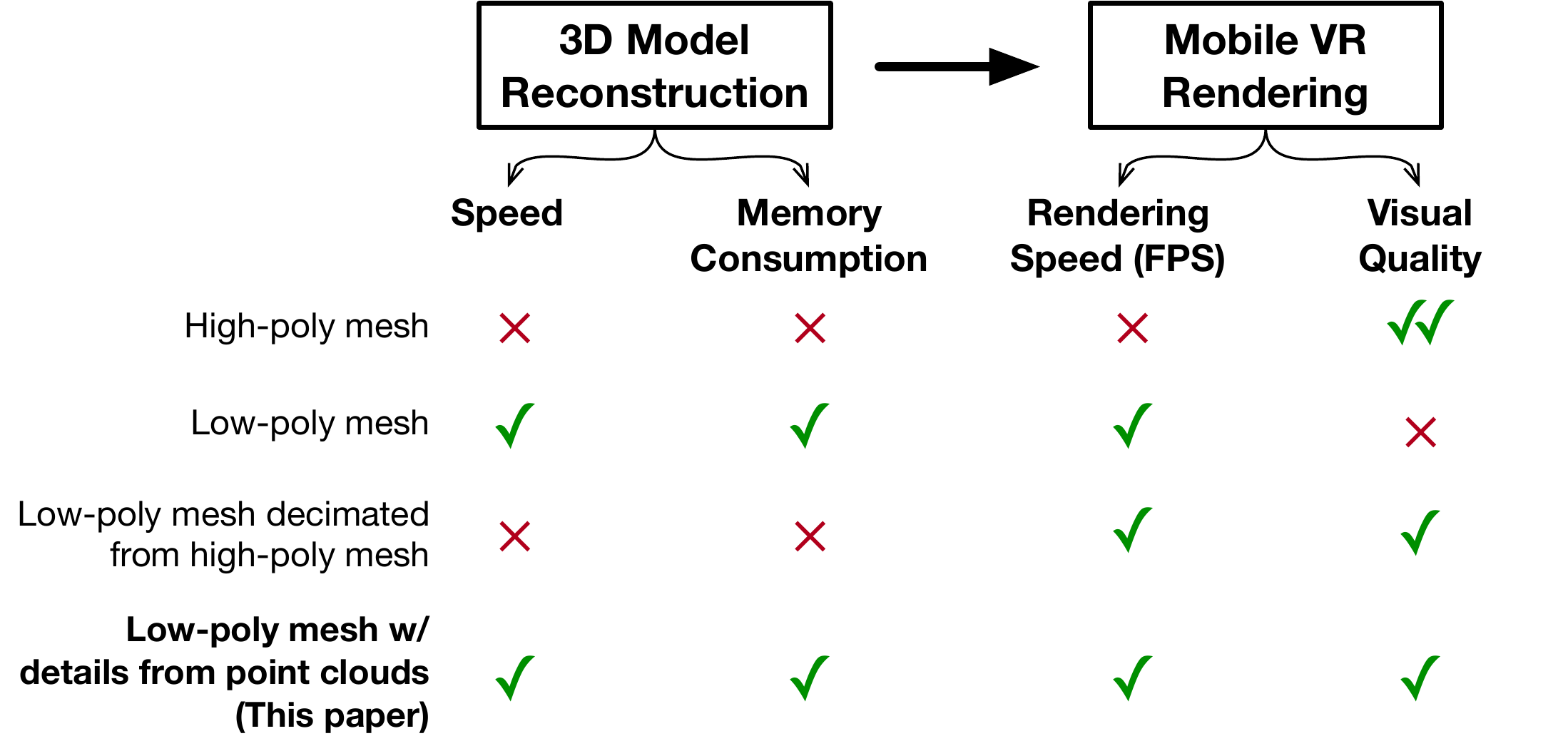}
    \caption{The design objective of our workflow is to simultaneously ensure high reconstruction efficiency (speed and RAM requirement) as well as high rendering speed and visual quality.}
    \label{fig:comparison}
\end{figure}

\paragraph{Real-Time Rendering Speed and Desirable Visual Quality} To faithfully represent the details of a scene a high-polygon dense mesh is preferred. Constructing meshes from point clouds is a well-studied problem in the literature~\cite{Kazhdan2006, Kazhdan2013}, and there are mature open-source and commercial software and libraries such as PCL~\cite{rusu2011point}, CGAL~\cite{CGAL}, MeshLab~\cite{MeshLab}, and Metashape~\cite{AgiSoftMetashape}.

However, dense meshes are hard to render in real-time, especially on mobile VR devices. Oculus Go suggests to keep the number of polygons under 100 thousand in order to achieve real-time rendering (60 FPS)~\cite{oculusperf}\footnote{Our experiments show that this is a conservative estimate. In general, a mesh with about 400 thousand faces renders in 60 FPS, but even a 400 thousand polygon mesh greatly loses visual details.}. On average, each room in the castle is scanned with 77 million points, which when simplified to under 100 thousand polygon loses much of the surface details that are critical for education, virtual tourism, and historical studies.

\paragraph{Reducing Reconstruction Overhead} To mitigate the loss of details while retaining high rendering speed, a common solution would be to decimate a high-polygon mesh to generate a low-polygon mesh and, optionally, transfer the details (e.g., texture and normal) from the high-polygon mesh to the low-polygon mesh. For each polygon in the low-polygon mesh, one could find its corresponding polygon in the high-polygon mesh and use the color and normal details there to generate the texture and normal for the low-polygon mesh. The decimation and transfer are readily supported in existing 3D modeling software such as Blender~\cite{Blender} and MeshLab~\cite{MeshLab}.


However, constructing high-polygon meshes (tens of millions of faces) and the texture map, either for direct rendering or for transferring details to low-polygon meshes, is extremely slow and requires high-end computers that are inaccessible to the broad archaeology and history communities. Using a high-end Intel Xeon server with 256 GB RAM and 2 Nvidia GTX 1080Ti GPUs, we estimate that reconstructing the entire castle would take three weeks using a typical workflow that consists of professional software such as Blender~\cite{Blender} and MeshLab~\cite{MeshLab} well-optimized for 3D modeling. In addition, since reconstruction processes huge amount of points and polygons, it consumes $\sim$20 GB RAM even for processing one single room. We thus had to reconstruct each room individually and stitch rooms later, further complicating the reconstruction process.





\paragraph{Summary} \Fig{fig:comparison} compares  the existing methods along four dimensions: the speed and memory consumption of reconstruction and the speed and visual quality of mobile VR rendering. Directly reconstructing and rendering a high-polygon mesh is slow in both reconstruction and rendering, albeit providing the best visual quality. In contrast, directly reconstructing and rendering a low-polygon mesh is efficient in both reconstruction and rendering, but suffers from low visual quality. Decimating a high-polygon mesh to a low-polygon mesh requires a high-polygon mesh to begin, and thus introduces high reconstruction overhead.

This paper proposes a reconstruction system that achieves high reconstruction efficiency (speed and RAM requirement) as well as high rendering speed and visual quality simultaneously. Specifically, our goal is to \textit{combine the best of both worlds}: delivering a reconstruction workflow that has a similar efficiency as using low-polygon meshes only while meeting (and even out-performing) the rendering quality of transferring details from a high-polygon mesh to a low-polygon mesh.

\section{Point-based Detail Transfer}
\label{sec:sys}

We propose a new reconstruction pipeline that directly transfer information from point cloud obtained from the laser scanner to low-polygon meshes without reconstructing high-polygon meshes altogether. This achieves both high reconstruction and rendering efficiency while preserving the visual details in the laser scan. We first provide an overview of the pipeline followed by describing our point-based detail transfer algorithm.

\subsection{Pipeline Overview}
\label{sec:sys:ov}

\Fig{fig:pipeline} shows the end-to-end reconstruction pipeline, which takes a laser-scanned colored point cloud as the input, and generates the mesh along with the texture and normal maps, which are then used for rendering on mobile VR platforms.

The input point cloud is first triangulated to generate a low-polygon mesh. We control the mesh to have a small amount of polygons to enable real-time rendering. For instance, when targeting Oculus Go, the mesh has at most 100 thousand polygons, which are about the threshold to deliver a 60 FPS rendering as suggested by Oculus~\cite{oculusperf}. We unwrap the mesh to generate the UV map, from which we bake the texture map and the normal map. Note that directly using these texture/normal maps for rendering would lead to low visual quality, because the texture and normal maps are generated from a low-polygon mesh. To increase the details in the texture/normal maps, one could increase the polygon count in a mesh at the cost of higher mesh reconstruction time and lower rendering speed. The left and middle panels in \Fig{fig:baking} below illustrate these two approaches.

Instead, our pipeline retains the same low-polygon UV map, but generates the textures and normals using information directly from the point cloud --- hence point-based detail transfer. The right panel in \Fig{fig:baking} illustrates this idea. We first map points from the point cloud to the low-polygon mesh (single triangle in this illustration); each pixel in the texture map is then interpolated using both the triangle vertices and the mapped points.

\begin{figure}[t]
    \centering
    \includegraphics[trim=0 0 35 0, clip, width=1\columnwidth]{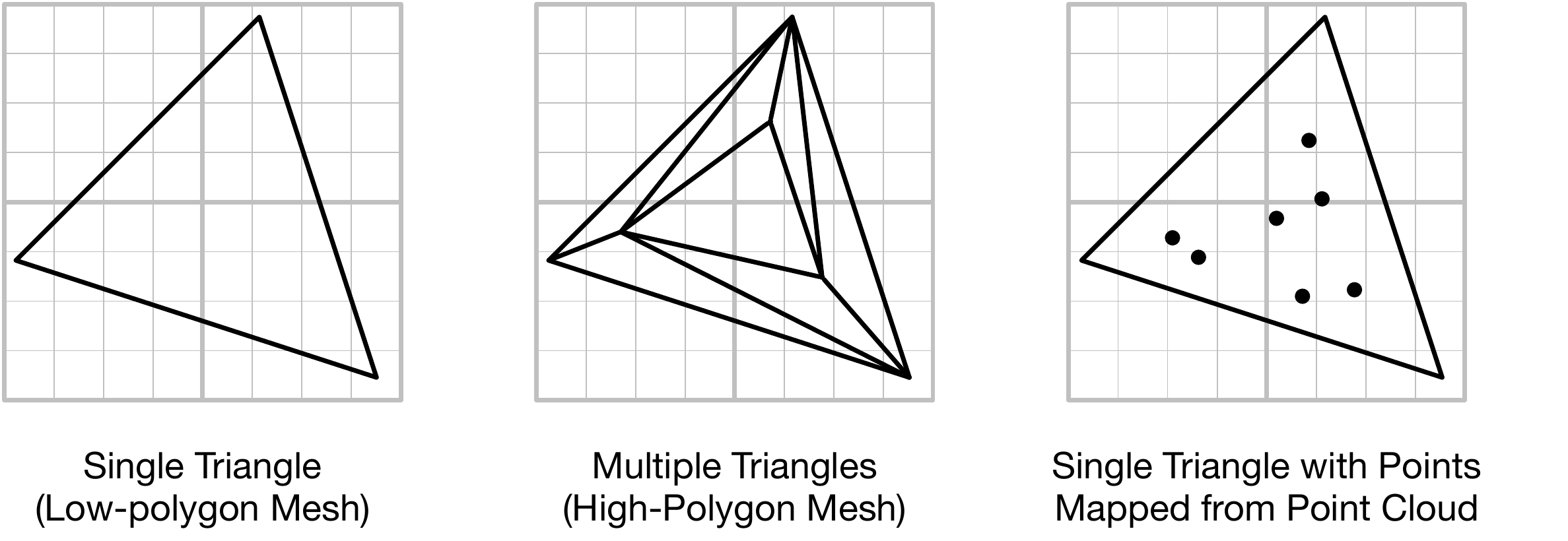}
    \caption{Three ways to generate texture (and normal). Left: interpolate from a low-polygon UV map (single triangle here); Middle: interpolate from a high-polygon UV map; Right (our method): map points from the point cloud to a low-polygon UV map, and then interpolate using the vertices and points.}
    \label{fig:baking}
\end{figure}

\begin{figure}[t]
    \centering
    \includegraphics[trim=0 0 0 0, clip, width=1\columnwidth]{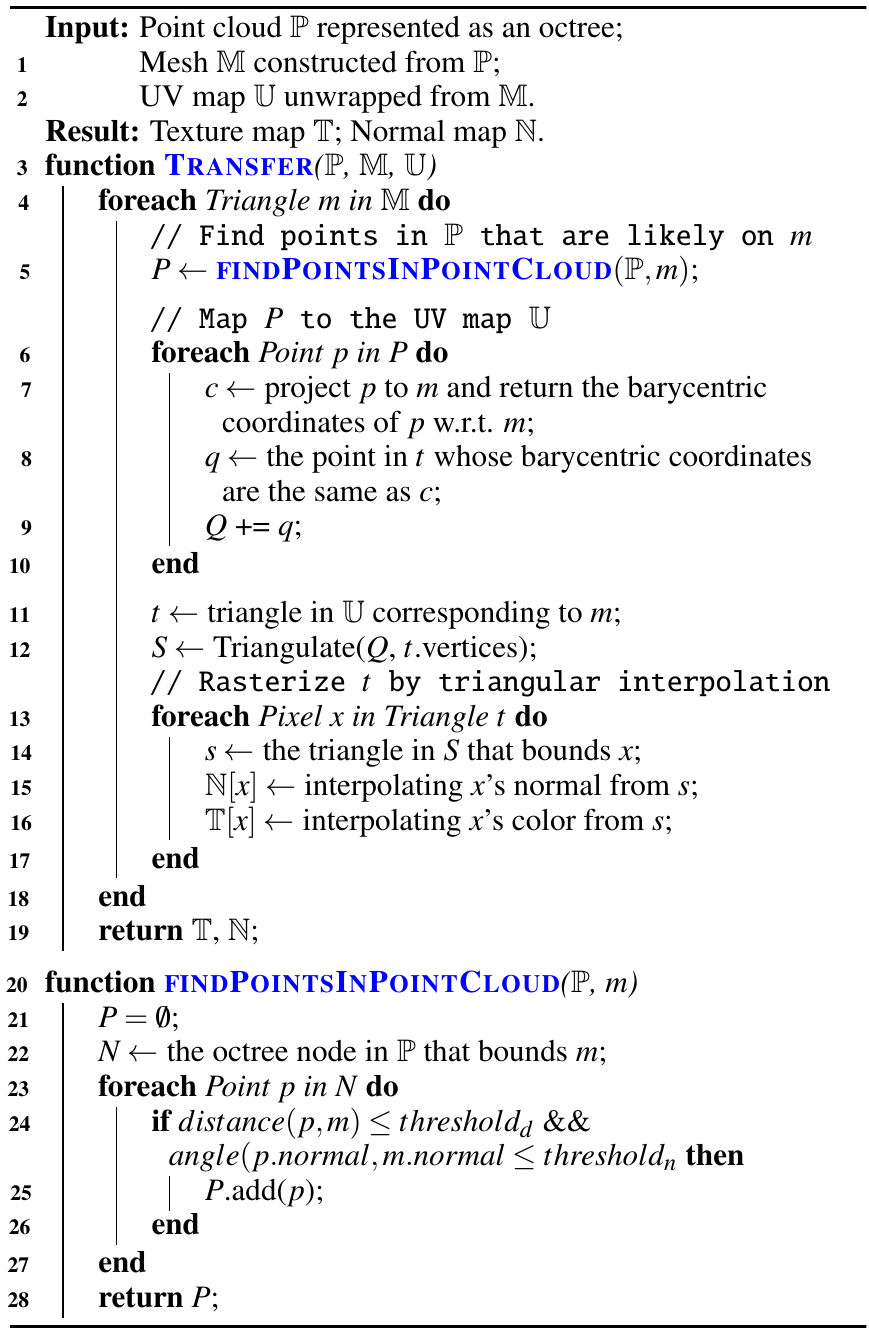}
    \caption{Pseudocode of our details transfer algorithm, which transfers details from a colored point cloud to a UV map, which is unwrapped from a mesh that is constructed from the point cloud.}
    \label{fig:algo}
\end{figure}

Compared to the low-polygon approach (left panel in \Fig{fig:baking}), our method provides more visual details because the textures/normals are generated using more information (triangle vertices + points vs. just triangle vertices). Compared to the high-polygon approach (middle panel in \Fig{fig:baking}), our method uses only a low-polygon mesh, and thus is faster.


\subsection{Detail Transfer Algorithm}
\label{sec:sys:algo}

\Fig{fig:algo} shows the pseudo-code of the point-based detail transfer algorithm. The goal is to map points from the point cloud to the (low-polygon) UV map so that we can generate the texture and normal maps with rich information to retain visual details. For each triangle in the mesh, we identify all the points in the point cloud that are likely on the triangle (Line 5). This is done by filtering points using their normals and distances to the triangle (Line 20 -- 28): points that are close to the triangle and whose normals are similar to the triangle normal are likely on the triangle and thus will be mapped to the UV map. The thresholds use for the distance and the angle between point and triangle are 4.0 and $120^{\circ}$, respectively. These parameters can be easily adjusted, and we empirically find that this setting works the best.

We map a point $p$ from the point cloud to the triangle in the UV map in a way that keeps the barycentric coordinates of $p$ constant with respect to the triangle (Line 6 -- 10), i.e., without changing the relative position of $p$ in the triangle. After point mapping, each triangle in the UV map could contain several points mapped from the point cloud. Each point is associated with its color and normal as in the point cloud. The right panel in \Fig{fig:baking} shows an example of a triangle with 7 points mapped from the point cloud.

With the transferred points, we can now generate the texture and normal maps (Line 11 -- 17). This is done by iterating over each triangle in the UV map and calculating the color and normal values of each pixel in the triangle. To leverage the transferred points in each triangle $T$, we first triangulate \{$T.V, T.P$\}, where $T.V$ and $T.P$ denote the vertices of $T$ and the transferred points in $T$, respectively. For each pixel $p$ in $T$, we find its bounding triangle from the triangulation results and use triangular interpolation to calculate the pixel's texture and normal. The same transfer process is applied to both the texture map and normal map, which along with the low-polygon mesh are fed into a VR rendering engine, e.g., Unity in our case, finishing our end-to-end pipeline.

%



\section{Evaluation}
\label{sec:eval}

We first describe our experimental methodology. We then show that the rendering quality of our system is competitive or better than today's typical reconstruction workflow, both objectively and subjectively, followed up by showing that our system is much faster and more resource efficient than today's workflow.

\subsection{Experimental Setup}
\label{sec:eval:exp}

\paragraph{Implementation Details}
We on average capture 8 scans of each room in the castle. We use FARO's Scene software to register point clouds of individual scans. The register point cloud becomes the input to our pipeline. We construct meshes using Meshlab~\cite{MeshLab}, a popular open-source mesh manipulation system, which implements the Screened Poisson Surface Reconstruction algorithm to build a triangulated mesh out of a point cloud~\cite{kazhdan2013screened}. We experimented with other mesh reconstruction tools, such as the commercial Agisoft Metashape software~\cite{AgiSoftMetashape}. The choice of reconstruction algorithm does not qualitatively change our conclusion. We use Meshlab in this paper because it allows us to build a completely open-source workflow that we will make available.

We use Blender~\cite{Blender} for UV unwrapping to generate a UV map, from which Blender is also used to generate the texture and normal map. We implement our point-based detail transfer in C++ based on the widely-used CGAL~\cite{CGAL}. The code is accelerated using OpenMP~\cite{openmp}.


\paragraph{Hardware} The reconstruction is done on a Dell Precision Workstation, which is equipped with a 4-core Intel Xeon W-2123 processor operating at 3.6 GHz, an Nvidia Quadro RTX4000 GPU, and 48 GB of RAM. This machine costs about \$2,100 to build, which we believe is a reasonable and representative machine specification for historical and archaeological researchers. We use Unity on Oculus Go for VR rendering. Oculus Go is a standalone VR headset that does not require tethering to a PC.

\paragraph{Dataset} We focus on three scenes in the Elmina castle: D11v3, NETower, and D31. They represent different indoor and outdoor scenes in Elmina Castle and are deemed representative by the archaeologists and historians in our team. For each scene, the camera viewpoint is chosen so that different geometries and textures are visible, e.g., including corners, windows, doors, uneven grounds, and walls with non-monotone colors.

For each scene, we generate a high-resolution mesh and a low-resolution mesh. \Tbl{tab:mesh} shows the number of faces in each mesh along with the corresponding point cloud size (measured in the number of points). We validate that the low-polygon meshes are able to be rendered on Oculus Go in 60 FPS.

\begin{table}[t]
\centering
  \caption{Point cloud size (number of points) and mesh size (number of polygons) of the three evaluated scenes.}
  \label{tab:mesh}
  \begin{tabular}{cccc}
    \toprule
    ~ & \textbf{D11v3}     & \textbf{NETower}   & \textbf{D31}       \\
    \midrule
    Point cloud & 28,368,767 & 8,652,472 & 31,431,682 \\
    High-polygon mesh & 4,352,296 & 6,857,430 & 3,690,788 \\
    Low-polygon mesh & 240,000   & 442,900   & 224,476  \\
    \bottomrule
  \end{tabular}
\end{table}

\paragraph{Baseline} We compare with three baselines (\Fig{fig:comparison}):
\begin{itemize}
	\item \hpm: constructing and rendering a high-polygon mesh.
	\item \lpm: constructing and rendering a low-polygon mesh.
	\item \trans: constructing a high-polygon mesh, simplifying/decimating it to a low-polygon mesh, and transferring color/normal details from the high-polygon mesh to the low-polygon mesh. The transfer is done in Blender, a professional 3D modeling software.
\end{itemize}

Meshlabs use the Octree to control the resolution of the point cloud. We use two different octree depths in the surface reconstruction algorithm to obtain the HPM and LPM. Our pipeline uses a low-polygon mesh throughout the workflow and never requires a high-polygon mesh.

It is worth noting again that our goal is \textit{not} to directly render points in the point cloud. As discussed in \Sect{sec:bck:reconstruct}, doing so would result in ``holes'' in the structure upon close examinations, which researchers in archaeology are strongly against. Therefore, we use \hpm as the ``ground-truth'' for quality comparison, as it constructs a high-quality, water-tight model.

We also have experimented with level-of-detail (LOD) rendering~\cite{xia1997adaptive} in Unity by manually providing meshes with different polygons and specifying when to use which mesh depending on the camera pose. While LOD is efficient when objects are viewed at distance, it still requires a high-polygon mesh, and thus does not provide real-time rendering, when viewing objects up close. In contrast, we address the speed issue by always rendering only a low-polygon mesh.

\paragraph{Metrics} We evaluate our system using three metrics:
\begin{itemize}
	\item Reconstruction time: the time it takes to generate the mesh and texture/normal maps required for VR rendering.
	\item Rendering speed: the FPS of rendering on Oculus Go.
	\item Visual quality: the mean-opinion-score in the HDR-VDP-2 metric~\cite{mantiuk2011hdr}, which is calibrated with user experience.
\end{itemize}

\subsection{Rendering Quality Comparison}
\label{sec:eval:render}

We show that our reconstruction system generates 3D models that, when rendered in mobile VR headsets, achieve higher visual quality than simply downsampling the mesh (i.e., \lpm) and match, sometimes outperform, the visual quality of transferring details from high-polygon mesh (i.e., \trans).

\paragraph{Objective Comparison} To quantitatively measure user experience, we export the rendering frames from the user perspective from Unity Player, and then compare our system with the baselines using the HDR-VDP-2 metric on the rendered frames. We use the frames from \hpm as the reference frames to calculate the HDR-VDP-2 metric for the other three systems. \Tbl{tab:quality} shows the results on the three scenes.

\begin{table}[t]
    \centering
    \caption{Rendering quality comparison in HDR-VDP-2.}
    \label{tab:quality}
    \begin{tabular}{c c c c}
        \toprule
        ~ & \textbf{D11v3} & \textbf{NETower} & \textbf{D31} \\
        \midrule
        \lpm & 54.01 & 51.34 & 50.50 \\
        \trans & 52.90 & 52.31 & 53.60 \\
        \textsc{Ours} & 55.58 & 52.24 & 54.00 \\
        \bottomrule
    \end{tabular}
\end{table}

Our reconstructed 3D models consistently outperform \lpm, indicating the benefit of transferring details. Compared to \trans, our reconstructed models have better visual quality in 2 out of the 3 scenes and come very close on the other ($<$ 0.1 in HDR-VDP-2 score). Considering that our reconstruction system delivers much faster speed (2.3 $\times$ speedup as we will show next), we believe our workflow provides a desirable design point.

\begin{figure*}[t]
    \centering
    \includegraphics[trim=0 0 0 0, clip, width=2\columnwidth]{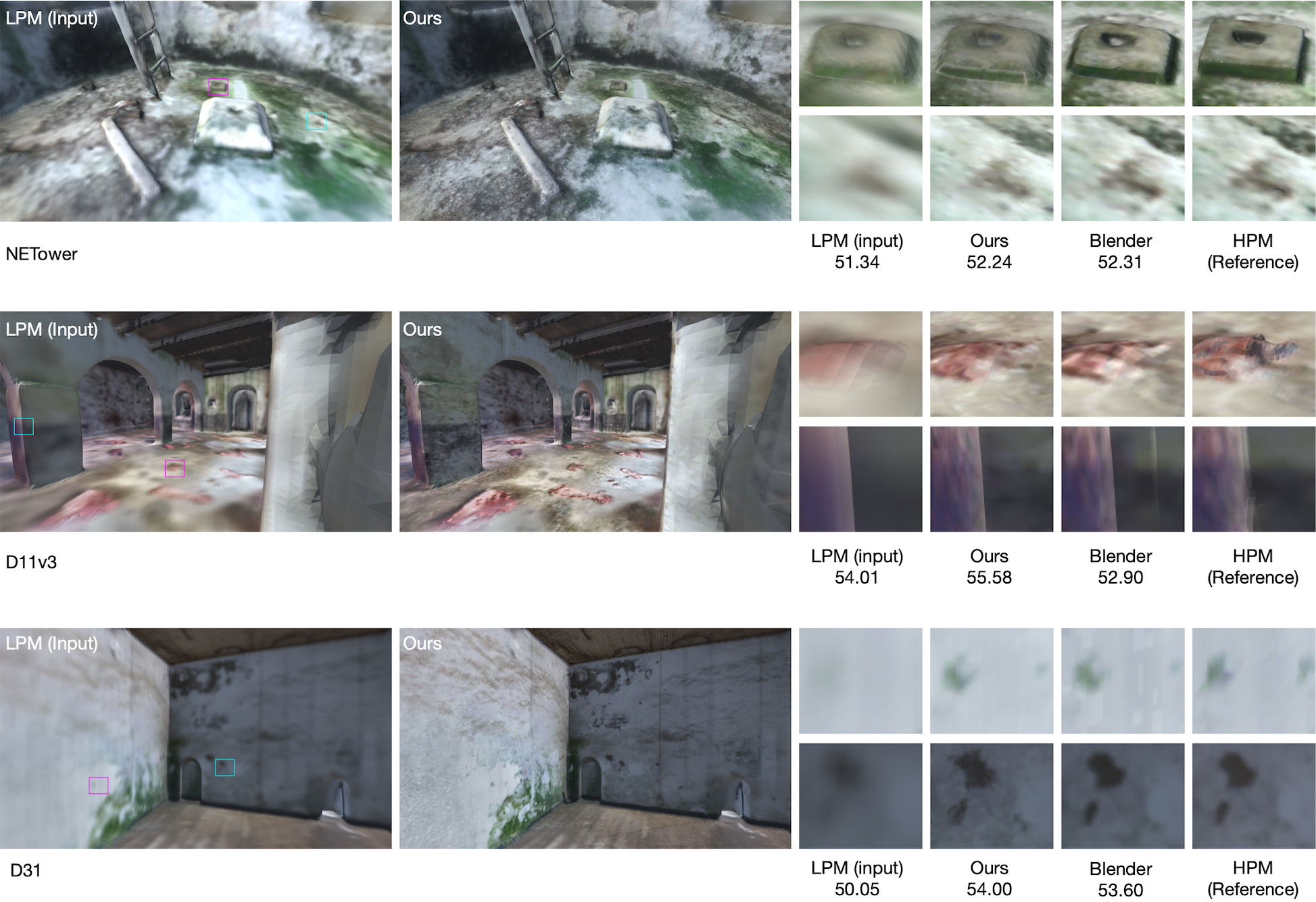}
    \caption{Visual results on three scenes in the castle and their HDR-VDP-2 scores. Our system generates frames that are visually competitive or better than \trans, a typical 3D reconstruction workflow.}
    \label{fig:subjectiveresults}
\end{figure*}

\paragraph{Subjective Comparison} Our rendering quality is subjectively competitive or better than that of \trans. \Fig{fig:subjectiveresults} shows the rendered frames of the three scenes under \lpm and ours; each frame has two zoom-in regions, for which we show the results of \lpm, ours, 
\trans, and \hpm.

Not surprisingly, \hpm consistently delivers the highest visual quality, due to the much larger meshes it constructs. Ours and \trans generate much better visual quality than \lpm. Comparing ours and \trans, our result is visually slightly worse in the NETower scene, but competitive or better in D11v3 and D31 scenes, which is also confirmed by the HDR-VDP-2 metric. In particular for D31 (the last scene), \trans generates significant blocking artifacts, which are invisible in our system.

\subsection{Efficiency Results}
\label{sec:eval:res}

\begin{figure}[t]
    \centering
    \subfloat[D11v3.]{
      \label{fig:d11v3_time}
      \includegraphics[height=1in]{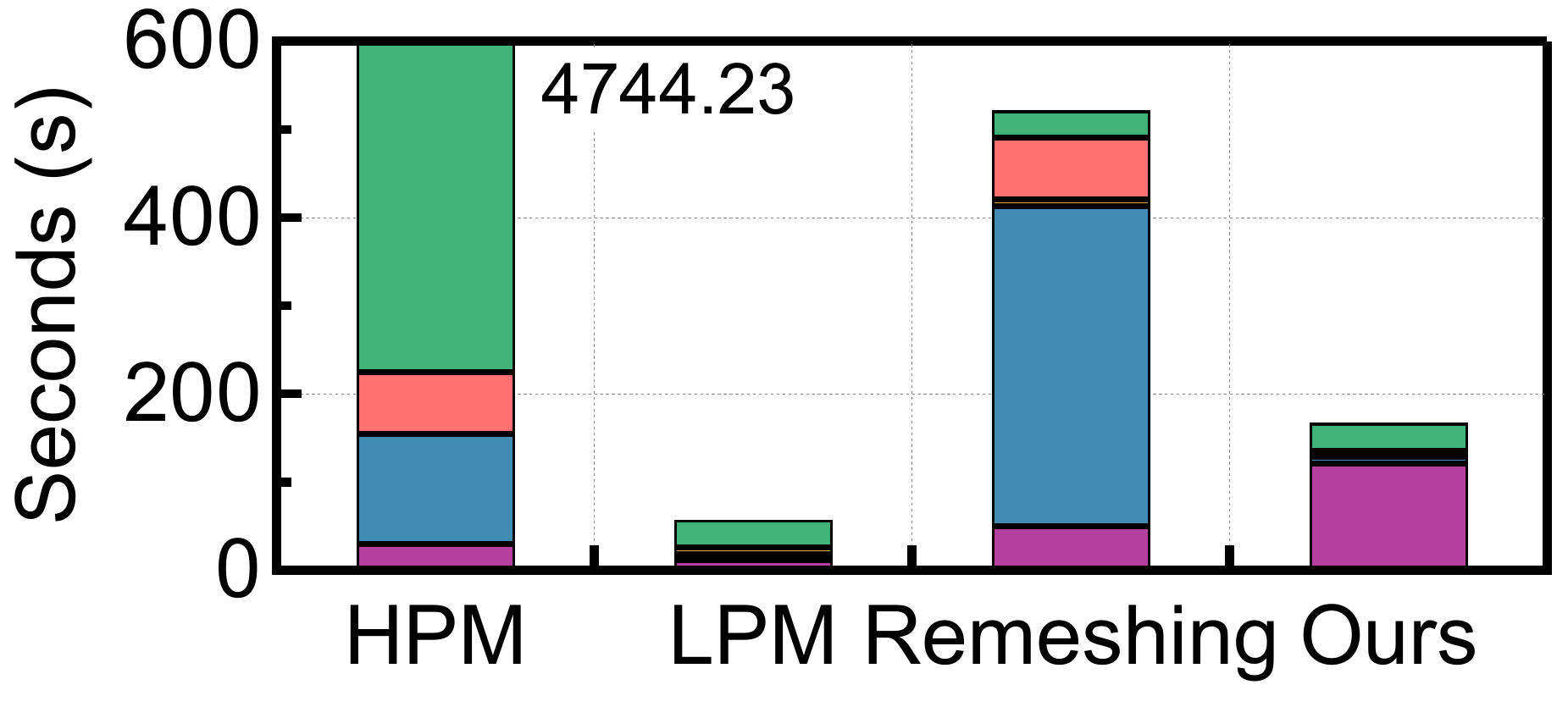}
    }
    \\
    \subfloat[NETower.]{
      \label{fig:ne_time}
      \includegraphics[height=1in]{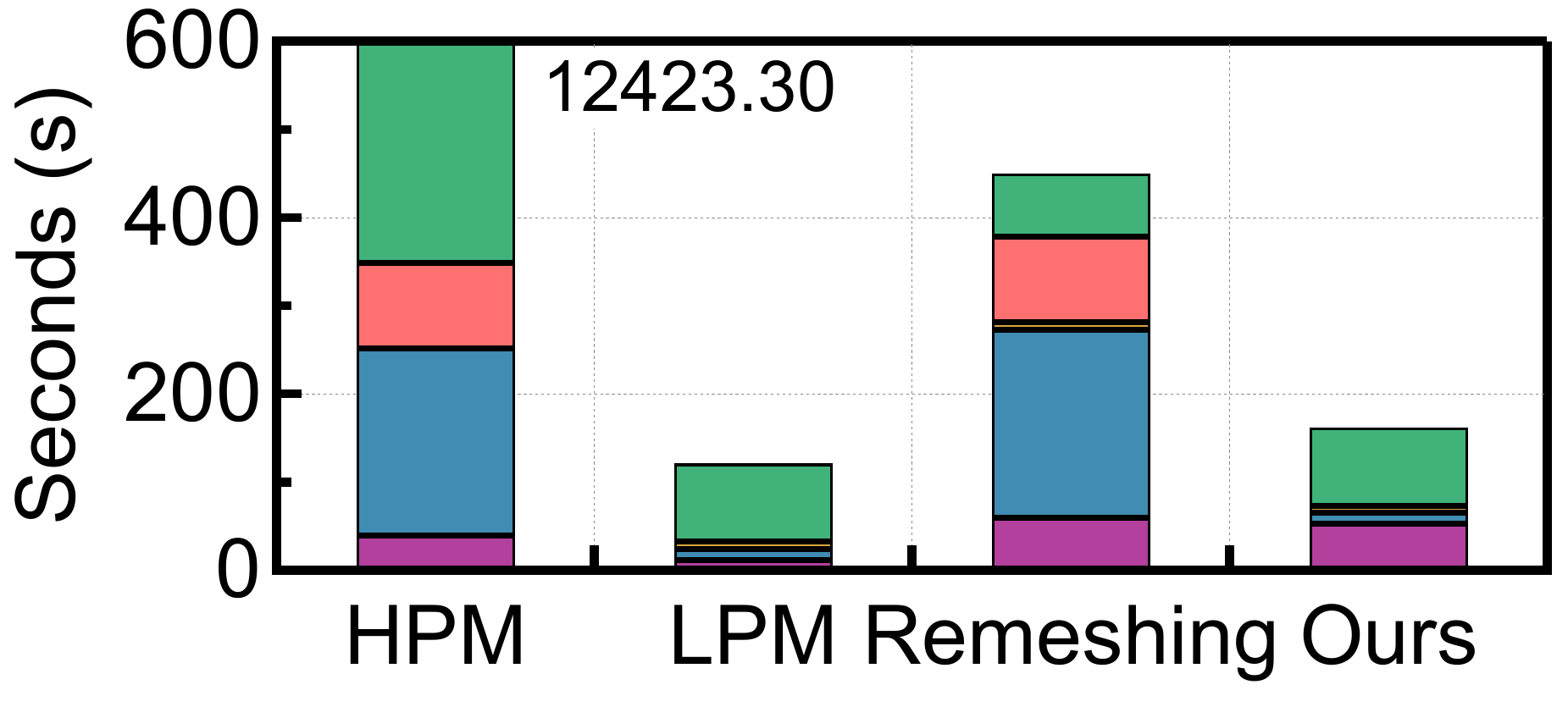}
    }
    \\
    \subfloat[D31.]{
      \label{fig:d31_time}
      \includegraphics[height=1in]{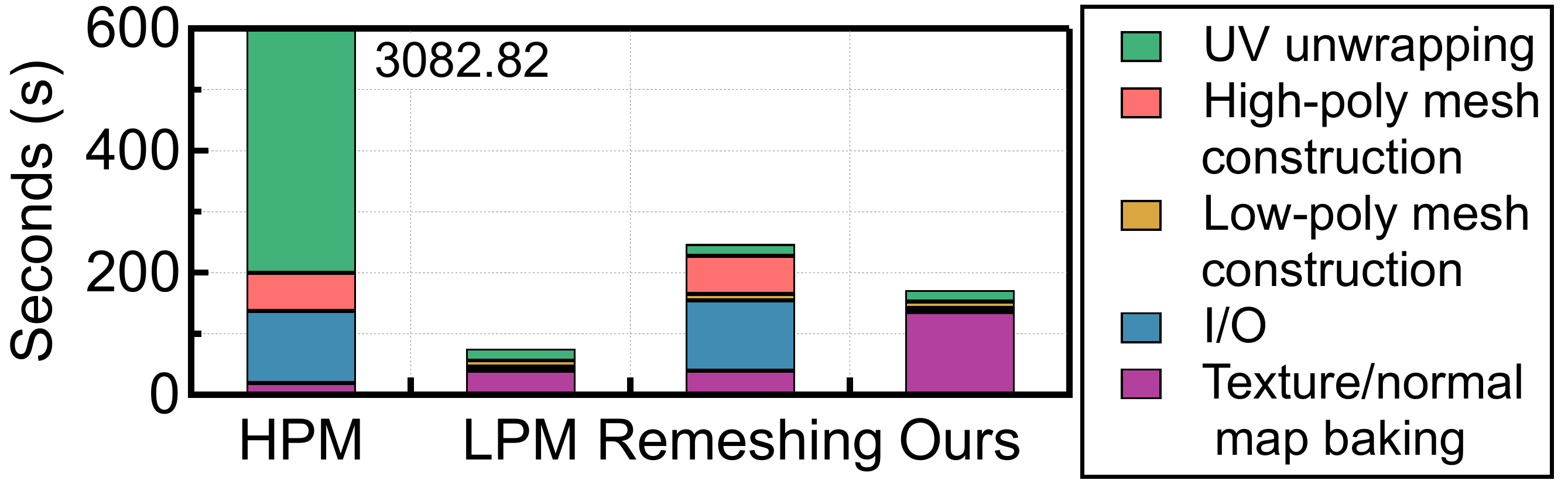}
    }
    \caption{End-to-end reconstruction time comparison. Our system is consistently faster (2.3 $\times$ speedup) than the typical \trans-based workflow, and is 33.8 $\times$ faster than \hpm.}
    \label{fig:time}
\end{figure}

\paragraph{Reconstruction Speed} We find that our workflow significantly reduces the end-to-end time to reconstruct a 3D model. \Fig{fig:time} compares the reconstruction time of our workflow with the three baselines across the three evaluated scenes. The stacked bar charts dissect the reconstruction time into five components: high-polygon mesh construction time, low-polygon mesh construction time, UV unwrapping time, I/O time (reading and writing meshes, texture/normal maps, and point clouds), and the texture/normal map generation (baking) time, which in our pipeline is dominated by the point-based transfer algorithm.

\hpm has the highest reconstruction time, which is dominated by UV unwrapping. \lpm has the lower reconstruction time by significantly reducing the UV unwrapping time, as it unwraps much smaller meshes. \lpm, however, has the lowest visual quality as we will show later.  \trans requires constructing a high-polygon mesh and introduces significant I/O overhead as it must manipulate high-polygon meshes throughout the workflow. Our point-based transfer is significantly faster than \trans by avoiding constructing and manipulating a high-polygon mesh. Across the three scenes, we achieve a 2.3 $\times$ and 33.8 $\times$ speedup over \trans and \hpm, respectively.

\begin{figure}[t]
    \centering
    \includegraphics[trim=0 0 0 0, clip, width=.9\columnwidth]{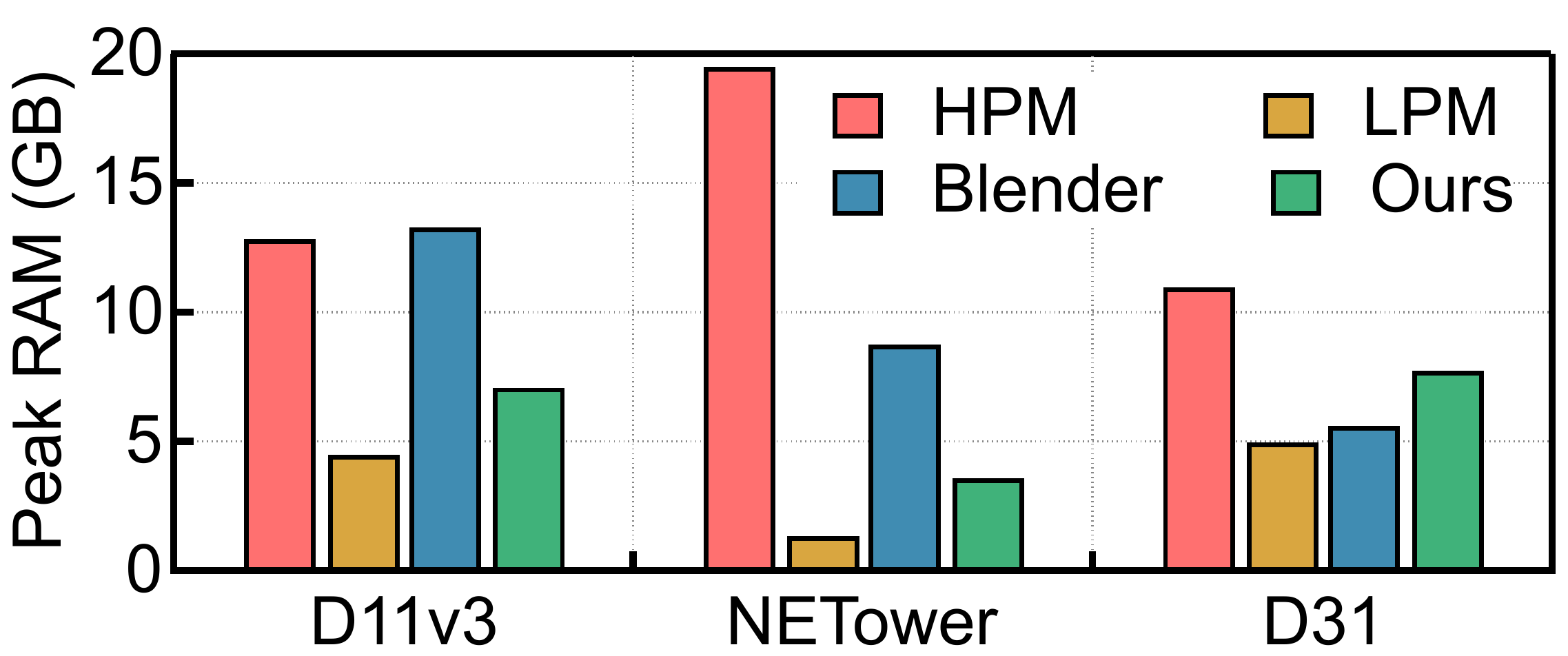}
    \caption{CPU RAM consumption comparison. Our system typically uses much less RAM than \trans and \hpm.}
    \label{fig:mem}
\end{figure}

\paragraph{RAM Consumption} \Fig{fig:mem} shows that our reconstruction workflow requires much less CPU RAM compared to \hpm and \trans since we avoid reconstructing and processing high-polygon meshes. On average, we reduce the peak RAM consumption by 51.8\% and 22.7\% compared to \hpm and \trans. Note that the peak memory consumption of our workflow is slightly higher on D31 compared to \trans. This is because D31 has the largest point cloud (\Tbl{tab:mesh}), on which our point transfer algorithm consumes high RAM.

\begin{figure}[t]
    \centering
    \includegraphics[trim=0 0 0 0, clip, width=.9\columnwidth]{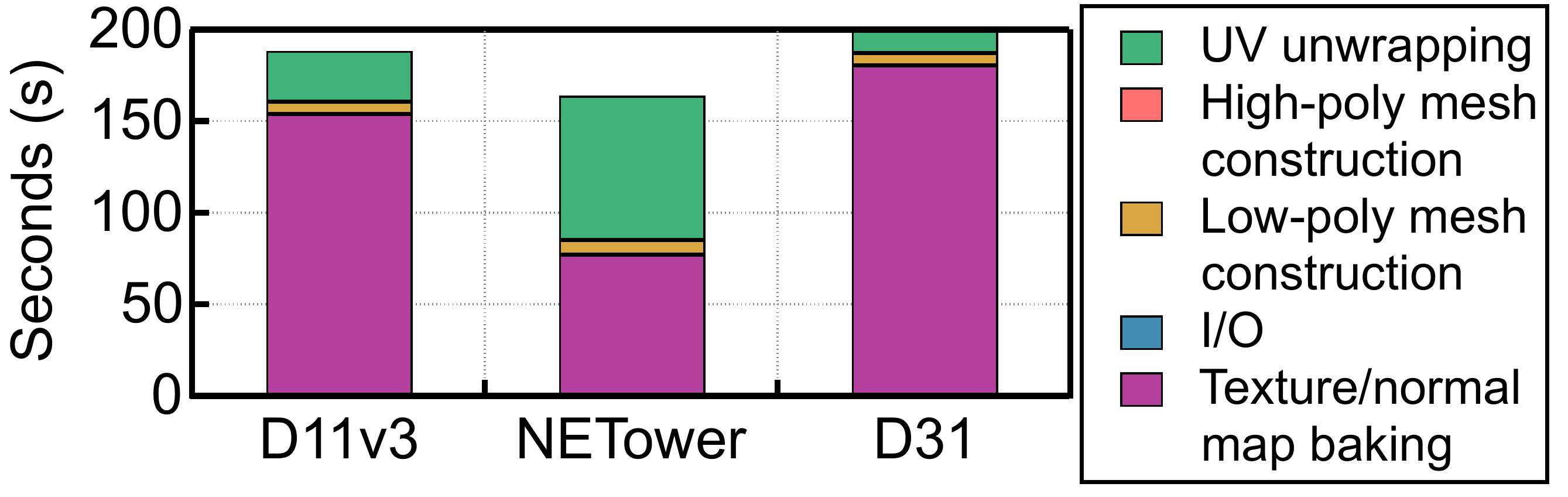}
    \caption{End-to-end reconstruction time on a mobile system.}
    \label{fig:mac_time}
\end{figure}

\paragraph{Rendering Speed} Using low-polygon meshes for VR rendering achieves 60 FPS, whereas using high-polygon meshes generally leads to an FPS below of 20. \Tbl{tab:fps} compares the FPS across the four workflows. Note that all except \hpm use low-polygon meshes and thus have the same 60 FPS.

\begin{table}[t]
    \centering
    \caption{FPS comparison.}
    \label{tab:fps}
    \begin{tabular}{c c c c}
        \toprule
        ~ & \textbf{D11v3} & \textbf{NETower} & \textbf{D31} \\
        \midrule
        \hpm     & 16    & 10      & 17  \\
        \textsc{Ours}/\lpm/\trans & 60    & 60      & 60 \\
        \bottomrule
    \end{tabular}
\end{table}

\paragraph{Evaluation on Low-end Systems} To further demonstrate that our workflow is resource-efficient and can be easily deployed even on a laptop, we also evaluate our end-to-end system on a 2015 MacBook Pro, which has  16~GB of CPU RAM and a four-core Intel i7-4870HQ processor running at 2.8~GHz.

\Fig{fig:mac_time} shows the reconstruction time on the laptop for the three scenes. While in generally it is slower on the laptop than on the workstation, the reconstruction time is consistently below 200 seconds for the three scenes, indicating that it is feasible to deploy our workflow on a mobile, low-end system. On average, running our reconstruction system on the mobile laptop is 10.6\% slower than on the Xeon workstation. Our system is also 33.1 $\times$ and 1.7 $\times$ faster than \hpm and \trans on the laptop.

\section{Related Work}
\label{sec:related}

\paragraph{Digitizing Cultural Heritage} Digital modeling and reconstruction has helped preserve and archive many cultural artifacts and heritage sites before. Perhaps the most well-known example is the efforts to digitally model the sculptures of Michelangelo by Levoy et. al.~\cite{levoy2000digital} and Rushmeier et. al.~\cite{bernardini2002building, rushmeier1998acquiring, rushmeier1997applying} two decades ago. The latter uses a structured-light scanner to capture range, color cameras to capture color, and a photometric stereo system to capture surface reflectance and normals. The former uses a laser-stripe scanner and a color camera to model the statues of Michelangelo and a time-of-flight rangefinder to scan building interiors. Our project relies on a time-of-flight ranging unit and a color camera (built in the FARO Focus3D X 130 scanner). As mentioned in \Sect{sec:bck:pc}, some rooms in the castle have to remain pitch-black, which excludes any device using visible light.

Compared to art objects, scanning and modeling large heritage sites tends to produce larger datasets with lower resolution, because heritage sites are much bigger. For instance, Michelangelo's statues are typically several meters tall; both of the prior modeling efforts produced geometry at (sub)-millimeter scale. In contrast, the footprint of the entire Elmina Castle is over \SI{8000}{\m\squared}; in our scans points are about 1.9 millimeters apart. Such a resolution is empirically sufficient for general entertainment, but higher resolution is preferred for historical education and research that require close-up examinations. It took us 500+ working hours to scan the castle at the current resolution.

\paragraph{Mobile VR Rendering} Our reconstruction targets mobile VR, which is resource-constrained and computationally weak. Much of the recent work focuses improving the VR/\ang{360} video rendering speed without sacrificing, using techniques ranging from memoization~\cite{boos2016flashback}, remote/offloaded rendering~\cite{meng2020coterie, lai2019furion, liu2018cutting}, specialized hardware support~\cite{sun2020energy, leng2019energy}.

Fundamentally, these techniques focus only on VR rendering without considering the inputs to the render: mesh, texture/normal maps. This paper optimizes the reconstruction phase, which generates low-complexity mesh but highly detailed texture and normal maps by directly transferring details from the point cloud.




\paragraph{Point-based Graphics} Our point-based detail transfer algorithm draws inspiration from classic point-based graphics/rendering ligerature~\cite{gross2011point, levoy1985use, rusinkiewicz2000qsplat, pfister2000surfels}, which skips meshes altogether and renders images from points. While our algorithm does not completely avoid reconstructing a mesh, it avoids reconstructing \textit{high-polygon} meshes by directly transferring details from points to low-polygon meshes in order to reduce the mesh reconstruction overhead.

In the future, it would be interesting to study if we could directly apply classic point-based rendering techniques by skipping mesh altogether. This is especially relevant as 1) point cloud acquisition devices (e.g., time-of-flight laser scanners such as the one we used for our project) become more accessible, and 2) modeling high-resolution heritage sites requires huge point clouds, for which mesh reconstruction is prohibitively expensive.

\section{Conclusion}
\label{sec:conc}

We propose a system that efficiently reconstructs large heritage sites from laser-scan point clouds while enabling real-time mobile VR rendering. We demonstrate the system on Elmina Castle in Ghana. We show that it is possible to deliver real-time rendering on mobile VR devices with high visual quality of a complicated architecture without requiring a high-polygon mesh. The key is to directly transfer details from the point cloud, which also significantly speeds up the reconstruction time and makes it more accessible for archaeologists and historians to reconstruct large heritage sites from huge point clouds using everyday computers.




\small
\bibliographystyle{plain}
\bibliography{refs}

\begin{thebibliography}{10}

\bibitem{bermuda100}
Bermuda 100.
\newblock \url{http://bermuda100.ucsd.edu/}.

\bibitem{cyark}
Cyark.
\newblock \url{https://www.cyark.org/}.

\bibitem{farox130}
Faro laser scanner focus3d x 130 manual.
\newblock \url{https://faro.app.box.com/s/r45cyjqengcts8vnh5kawemgsvdfxt81}.

\bibitem{oculusperf}
Oculus testing and performance analysis.
\newblock \url{https://developer.oculus.com/documentation/unity/unity-perf/}.

\bibitem{openheritage}
Open heritage 3d.
\newblock \url{https://openheritage3d.org/}.

\bibitem{openmp}
Openmp.
\newblock \url{https://www.openmp.org/}.

\bibitem{AgiSoftMetashape}
{Agisoft, LLC.}
\newblock Agisoft metashape.
\newblock \url{https://www.agisoft.com}.

\bibitem{bernardini2002building}
Fausto Bernardini, Holly Rushmeier, Ioana~M Martin, Joshua Mittleman, and
  Gabriel Taubin.
\newblock Building a digital model of michelangelo's florentine pieta.
\newblock {\em IEEE Computer Graphics and Applications}, 22(1):59--67, 2002.

\bibitem{bonatto2016explorations}
Daniele Bonatto, S{\'e}gol{\`e}ne Rogge, Arnaud Schenkel, Rudy Ercek, and
  Gauthier Lafruit.
\newblock Explorations for real-time point cloud rendering of natural scenes in
  virtual reality.
\newblock In {\em 2016 International Conference on 3D Imaging (IC3D)}, pages
  1--7. IEEE, 2016.

\bibitem{boos2016flashback}
Kevin Boos, David Chu, and Eduardo Cuervo.
\newblock Flashback: Immersive virtual reality on mobile devices via rendering
  memoization.
\newblock In {\em Proceedings of the 14th Annual International Conference on
  Mobile Systems, Applications, and Services (MobiSys'16)}, pages 291--304.
  ACM, 2016.

\bibitem{champion2003applying}
Erik Champion.
\newblock Applying game design theory to virtual heritage environments.
\newblock In {\em Proceedings of the 1st international conference on Computer
  graphics and interactive techniques in Australasia and South East Asia},
  pages 273--274, 2003.

\bibitem{MeshLab}
Paolo Cignoni, Marco Callieri, Massimiliano Corsini, Matteo Dellepiane, Fabio
  Ganovelli, and Guido Ranzuglia.
\newblock {MeshLab: an Open-Source Mesh Processing Tool}.
\newblock In Vittorio Scarano, Rosario~De Chiara, and Ugo Erra, editors, {\em
  Eurographics Italian Chapter Conference}. The Eurographics Association, 2008.

\bibitem{cohen1998appearance}
Jonathan Cohen, Marc Olano, and Dinesh Manocha.
\newblock Appearance-preserving simplification.
\newblock In {\em Proceedings of the 25th annual conference on Computer
  graphics and interactive techniques}, pages 115--122, 1998.

\bibitem{Blender}
Blender~Online Community.
\newblock Blender - a 3d modelling and rendering package, 2018.

\bibitem{garland1997surface}
Michael Garland and Paul~S Heckbert.
\newblock Surface simplification using quadric error metrics.
\newblock In {\em Proceedings of the 24th annual conference on Computer
  graphics and interactive techniques}, pages 209--216, 1997.

\bibitem{gee2003video}
James~Paul Gee.
\newblock What video games have to teach us about learning and literacy.
\newblock {\em Computers in Entertainment (CIE)}, 1(1):20--20, 2003.

\bibitem{gross2011point}
Markus Gross and Hanspeter Pfister.
\newblock {\em Point-based graphics}.
\newblock Elsevier, 2011.

\bibitem{kapell2013playing}
Matthew~Wilhelm Kapell and Andrew~BR Elliott.
\newblock {\em Playing with the past: Digital games and the simulation of
  history}.
\newblock Bloomsbury Publishing USA, 2013.

\bibitem{Kazhdan2006}
Michael Kazhdan, Matthew Bolitho, and Hugues Hoppe.
\newblock Poisson surface reconstruction.
\newblock In {\em Proceedings of the Fourth Eurographics Symposium on Geometry
  Processing}, SGP ’06, page 61–70, Goslar, DEU, 2006. Eurographics
  Association.

\bibitem{Kazhdan2013}
Michael Kazhdan and Hugues Hoppe.
\newblock Screened poisson surface reconstruction.
\newblock {\em ACM Trans. Graph.}, 32(3), July 2013.

\bibitem{kazhdan2013screened}
Michael Kazhdan and Hugues Hoppe.
\newblock Screened poisson surface reconstruction.
\newblock {\em ACM Transactions on Graphics (ToG)}, 32(3):1--13, 2013.

\bibitem{lai2019furion}
Zeqi Lai, Y~Charlie Hu, Yong Cui, Linhui Sun, Ningwei Dai, and Hung-Sheng Lee.
\newblock Furion: Engineering high-quality immersive virtual reality on today's
  mobile devices.
\newblock {\em IEEE Transactions on Mobile Computing}, 2019.

\bibitem{leng2019energy}
Yue Leng, Chi-Chun Chen, Qiuyue Sun, Jian Huang, and Yuhao Zhu.
\newblock Energy-efficient video processing for virtual reality.
\newblock In {\em Proceedings of the 46th International Symposium on Computer
  Architecture}, pages 91--103, 2019.

\bibitem{levoy2000digital}
Marc Levoy, Kari Pulli, Brian Curless, Szymon Rusinkiewicz, David Koller, Lucas
  Pereira, Matt Ginzton, Sean Anderson, James Davis, Jeremy Ginsberg, et~al.
\newblock The digital michelangelo project: 3d scanning of large statues.
\newblock In {\em Proceedings of the 27th annual conference on Computer
  graphics and interactive techniques}, pages 131--144, 2000.

\bibitem{levoy1985use}
Marc Levoy and Turner Whitted.
\newblock {\em The use of points as a display primitive}.
\newblock Citeseer, 1985.

\bibitem{liu2018cutting}
Luyang Liu, Ruiguang Zhong, Wuyang Zhang, Yunxin Liu, Jiansong Zhang, Lintao
  Zhang, and Marco Gruteser.
\newblock Cutting the cord: Designing a high-quality untethered vr system with
  low latency remote rendering.
\newblock In {\em Proceedings of the 16th Annual International Conference on
  Mobile Systems, Applications, and Services}, pages 68--80, 2018.

\bibitem{mantiuk2011hdr}
Rafat Mantiuk, Kil~Joong Kim, Allan~G Rempel, and Wolfgang Heidrich.
\newblock Hdr-vdp-2: A calibrated visual metric for visibility and quality
  predictions in all luminance conditions.
\newblock {\em ACM Transactions on graphics (TOG)}, 30(4):1--14, 2011.

\bibitem{meng2020coterie}
Jiayi Meng, Sibendu Paul, and Y~Charlie Hu.
\newblock Coterie: Exploiting frame similarity to enable high-quality
  multiplayer vr on commodity mobile devices.
\newblock In {\em Proceedings of the Twenty-Fifth International Conference on
  Architectural Support for Programming Languages and Operating Systems}, pages
  923--937, 2020.

\bibitem{pfister2000surfels}
Hanspeter Pfister, Matthias Zwicker, Jeroen Van~Baar, and Markus Gross.
\newblock Surfels: Surface elements as rendering primitives.
\newblock In {\em Proceedings of the 27th annual conference on Computer
  graphics and interactive techniques}, pages 335--342, 2000.

\bibitem{rushmeier1998acquiring}
Holly Rushmeier, Fausto Bernardini, Joshua Mittleman, and Gabriel Taubin.
\newblock Acquiring input for rendering at appropriate levels of detail:
  Digitizing a pieta.
\newblock In {\em Rendering Techniques’ 98}, pages 81--92. Springer, 1998.

\bibitem{rushmeier1997applying}
Holly Rushmeier, Gabriel Taubin, and Andr{\'e} Gu{\'e}ziec.
\newblock Applying shape from lighting variation to bump map capture.
\newblock In {\em Rendering Techniques’ 97}, pages 35--44. Springer, 1997.

\bibitem{rusinkiewicz2000qsplat}
Szymon Rusinkiewicz and Marc Levoy.
\newblock Qsplat: A multiresolution point rendering system for large meshes.
\newblock In {\em Proceedings of the 27th annual conference on Computer
  graphics and interactive techniques}, pages 343--352, 2000.

\bibitem{rusu2011point}
Radu~B Rusu and S~Cousins.
\newblock Point cloud library (pcl).
\newblock In {\em 2011 IEEE International Conference on Robotics and
  Automation}, pages 1--4, 2011.

\bibitem{schutz2019real}
Markus Sch{\"u}tz, Katharina Kr{\"o}sl, and Michael Wimmer.
\newblock Real-time continuous level of detail rendering of point clouds.
\newblock In {\em 2019 IEEE Conference on Virtual Reality and 3D User
  Interfaces (VR)}, pages 103--110. IEEE, 2019.

\bibitem{squire2011video}
Kurt Squire.
\newblock Video games and learning.
\newblock {\em Teaching and participatory culture in the digital age}, 2011.

\bibitem{sun2020energy}
Qiuyue Sun, Amir Taherin, Yawo Siatitse, and Yuhao Zhu.
\newblock Energy-efficient 360-degree video rendering on fpga via
  algorithm-architecture co-design.
\newblock In {\em The 2020 ACM/SIGDA International Symposium on
  Field-Programmable Gate Arrays}, pages 97--103, 2020.

\bibitem{CGAL}
{The CGAL Project}.
\newblock {\em {CGAL} User and Reference Manual}.
\newblock {CGAL Editorial Board}, {5.0} edition, 2019.

\bibitem{xia1997adaptive}
Julie~C. Xia, Jihad El-Sana, and Amitabh Varshney.
\newblock Adaptive real-time level-of-detail based rendering for polygonal
  models.
\newblock {\em IEEE Transactions on Visualization and Computer graphics},
  3(2):171--183, 1997.

\end{thebibliography}


\begin{biography}

Sifan Ye is a graduate student at Stanford University. He obtained his BS in Computer Science from University of Rochester. His work was done in part while at University of Rochester.

Ting Wu is a software engineer at eBay. She obtained Bachelor's degree in Vehicle Engineering from Tongji University (2017) and Master's degree in Computer Science from University of Rochester (2020). Her work was done entirely while at University of Rochester.

Michael Jarvis is an Associate Professor of History, Director of Digital Media Studies Program, and Director of Smiths Island Archaeology Project at the University of Rochester. He works onEarly American, Atlantic, Maritime, Public and Digital history and historical archaeology. He obtained his Ph.D. from College of William and Mary (1998).

Yuhao Zhu received his BS in Computer Science from Beihang University (2010) 
and his Ph.D. in Electrical and Computer Engineering from The University of Texas at Austin (2017). He is an Assistant Professor of Computer Science at University of Rochester. His work focuses on applications, algorithms, and systems for visual computing.
\end{biography}

\end{document}